\documentclass[aps,prc,twocolumn,showpacs]{revtex4}
\usepackage[dvips]{graphicx}
\usepackage{dcolumn}
\usepackage{bm}

\begin{document}

\title{Order parameter fluctuations in percolation:
Application to nuclear multifragmentation}

\author{Janusz Brzychczyk} 

\email{ufjanusz@if.uj.edu.pl}

\affiliation{M. Smoluchowski Institute of Physics, Jagiellonian 
University, Reymonta 4, 30-059 Krak\'ow, Poland}

\date{\today}

\begin{abstract}
Order parameter fluctuations (the largest cluster size distribution)
are studied within a three-dimensional bond percolation model on small
lattices. Cumulant ratios measuring the fluctuations exhibit distinct
features near the percolation transition (pseudocritical point),
providing a method for its identification. The location of
the critical point in the continuous limit can be estimated without
variation of the system size.
This method is remarkably insensitive to finite-size effects
and may be applied even for a very small system.
The possibility of using various measurable quantities
for sorting events makes the procedure useful in studying
clusterization phenomena, in particular nuclear multifragmentation.
Finite-size scaling and $\Delta$-scaling relations are examined.
The model shows inconsistency with some of the $\Delta$-scaling
expectations. The role of surface effects is evaluated by comparing
results for free and periodic boundary conditions.
\end{abstract}

\pacs{24.60.Ky, 25.70.Pq, 05.70.Jk, 64.60.Ak}

\maketitle


\section{Introduction}

The main motivation of nuclear multifragmentation studies
is probing a liquid-gas coexistence region in the phase diagram
of nuclear matter \cite{Siemens,Jaq,good}.
Many works deduce the occurrence of
a first- or second-order phase transition
\cite{Finn,gil,poch,ago,d-p,gupta,kleine,fis,nat,sri,ma}.
Although both transition types can be expected,
unambiguous identification is difficult due to strong finite-size
effects in systems with a small number of constituents.
In such systems, for example, a first-order phase transition may mimic
critical behavior \cite{mim,car}. On the other hand, nuclear
multifragmentation induced by high energy collisions
shows striking similarities to percolation processes which are known 
to contain a second-order phase transition
(critical behavior) \cite{kleine,b1,b2,c1,c2,kreutz,li,hyp}.
Percolation-based models seem to be also successful in describing
fragmentation of atomic clusters \cite{far,gobet}.
These observations lead to formulation of the hypothesis that percolation
could be a universal fragmentation mechanism for simple fluids \cite{hyp}.
For better recognizing critical-like behavior observed in fragmentation
processes, simultaneous application of various complementary methods is
necessary. Percolation models are often used to construct or verify
procedures tracing critical behavior in fragmenting systems
\cite{b1,b2,c1,c2,el,ell,d-e,uf}. They provide
a simple tool for studying universal aspects of the critical behavior
and the role of finite-size effects.

In the framework of a percolation model we have examined the largest cluster
size distribution. The size of the largest cluster, as an order parameter
in aggregation scenarios of the fragment production, is of particular
interest in phase transition studies  \cite{d-e,uf}.
The limiting forms of the distribution for normal phases are predicted
by classical limit theorems for random variables \cite{uf,d-n,baz1,baz2}.
At a second-order phase transition the system is highly correlated
with fluctuations occurring on all length scales.
Properties of the order parameter close to the critical point
can be studied with the renormalization group and finite-size scaling
approaches \cite{d-e,uf,d-n,baz2,sta,binder}. Botet and P{\l}oszajczak
have proposed to identify the second-order critical behavior
in finite systems by examining universal features of the order parameter
fluctuations with a $\Delta$-scaling method \cite{d-p,d-e,uf,d-n}.
The method has been applied to several models and nuclear fragmentation
data \cite{d-p,ma,car,d-e,uf,fra,frank,frankland,gul}. We will confront
$\Delta$-scaling predictions with percolation results. 

In order to compare theoretical predictions with fragmentation
data all experimental conditions should be carefully considered.
The bulk behavior of the order parameter can be significantly
modified in small fragmenting systems by finite-size and boundary effects.
The control parameter is usually not well measured and must
be substituted by other measurable quantities in sorting events,
leading to additional modifications.
We aim to evaluate the significance of such effects in the present work.

The calculations were performed with a three-dimensional bond
percolation model on the simple cubic lattices \cite{sta}.
A Monte Carlo procedure with the Hoshen-Kopelman cluster labeling
algorithm was employed to generate events
for a distribution of the bond probability, $p$, being the control parameter.
The lattices of size $N=L^3$ with $L=3,4,5,6$ correspond to the range of
system sizes available in nuclear reactions. Free boundary conditions were
applied to account for the presence of surface in real systems. To
evaluate the role and importance of finite-size and boundary effects the
calculations were extended to include larger systems and periodic boundary
conditions.

We study low-order cumulants (cumulant ratios) as the mean, variance,
skewness and kurtosis of the largest cluster size distribution.
These standard statistical measures contain the most significant
information, providing a robust identification of the percolation transition.
In order to place our results in a wider context,
we will briefly recall in the next section
some signals of criticality that are frequently tested in
fragmentation studies.

\section{Percolation transition in small systems}

In the strictest sense, the phase transition occurs in the continuous
limit $N\rightarrow\infty$. Then, below the percolation threshold
$p<p_{c}\simeq 0.2488$, only finite clusters are present.
When $p>p_{c}$ there exists an infinite cluster spanning the whole
lattice. The fraction of sites belonging to this cluster is the order
parameter. In finite systems the transition is smoothed.
The probability that at least one cluster connects the bottom and the top
lattice planes changes gradually as illustrated in Fig. 1(a).
\begin{figure}[ht]
\includegraphics[width=3.375in]{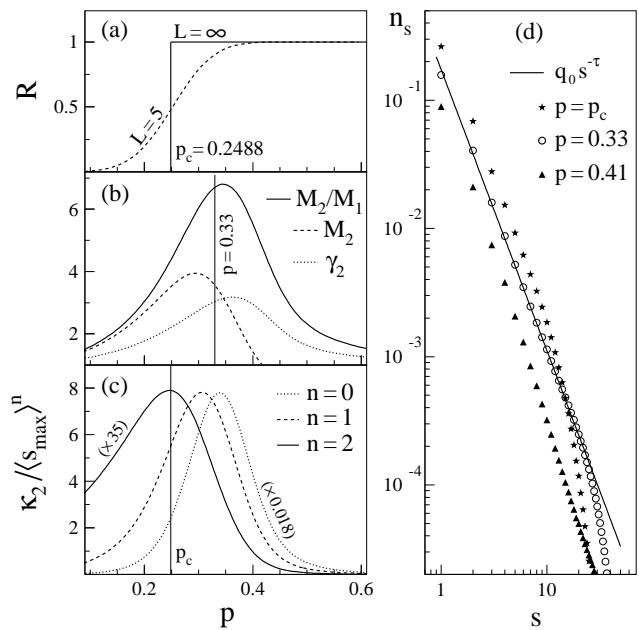}
\caption{Bond percolation on a lattice of linear size $L=5$
with free boundary conditions.
Plotted as a function of the bond probability:
(a) the probability that a lattice is spanned,
(b) moments of the average distribution of cluster sizes,
(c) the variance of the largest cluster size distribution with various
normalizations.
(d) The average distribution of cluster sizes.}
\label{fig:1}
\end{figure}
A natural way to locate the transition is to
examine quantities which diverge at the critical point in the continuous
limit. For finite systems, the divergence is replaced by a maximum
located near $p_c$. This can be seen in Fig. 1(b) for
the second moment of the average distribution of cluster sizes, $M_2$,
the mean cluster size, $M_{2}/M_{1}$,
and the reduced variance, $\gamma_{2}\equiv M_{2}M_{0}/M_{1}^2$ \cite{c2}.
$M_{k}$ denotes the $k$-th moment of the cluster size distribution,
$M_{k}=\sum_{s} s^{k}n_{s}$, where $n_s$ is
the average number of clusters of size $s$
normalized to the system size (the largest cluster excluded).
The mean cluster size is the analog of the susceptibility and the location
of its maximum value defines the pseudocritical point.
Another frequently used method is a power-law fit to the fragment size
distribution \cite{Finn,ell}.
In our example the best fit is obtained for
$p=0.33$ as shown in Fig. 1(d). The fragment size distribution
follows in some range the asymptotic behavior $n_{s}=q_{0}s^{-\tau}$,
where the Fisher exponent $\tau=2.189$ \cite{exp}. The normalization
constant $q_0=1/\sum_{s=1}^{\infty} s^{(1-\tau)}$ is taken as $0.173$
from the summation computed up to $s=10^{9}$.
Another example is the maximum fluctuation of the largest cluster size.
The size of the largest cluster, $s_{max}$,
plays the role of an extensive order parameter.
Its fluctuations are usually measured by the variance, $\kappa_{2}$,
or the normalized variance, $\kappa_{2}/\langle s_{max}\rangle$,
of the probability distribution $P(s_{max})$ \cite{c2,ell,hyp,dorso}.
The two quantities are peaked in the transition region as shown
in Fig. 1(c) by the dotted and dashed lines.
When using the normalization $\kappa_{2}/\langle s_{max}\rangle ^2$,
the maximum is located remarkably close to the critical point.

The above examples show that various investigated signals appear
at different positions and in most cases they are
shifted from the critical point toward the ordered phase region.
In small systems the shifts may be significant and should be
taken into account in a criticality analysis. The location of
the critical region in nuclear multifragmentation is
often deduced from a power-law fit to the fragment size distribution.
This location, appearing near the pseudocritical point,
would correspond to a temperature $T_{pc}$ distinctly lower than
the true critical temperature $T_{c}$.
For example, converting the bond probability to the temperature
with the prescription of Ref. \cite{li,babo},
one obtains $T_{pc}/T_{c}=0.64, 0.73$ and 0.78 for
$N=27,64$ and 125, respectively.

The location of the true critical point, $p_c$, is of particular
interest. According to the finite-size scaling
the position of a signal, $p(L)$, is expected to converge
to $p_{c}$ with increasing linear lattice size $L$ as
\begin{equation}
p(L)-p_{c}\propto L^{-1/\nu},
\end{equation}
where $\nu$ is the correlation length exponent. Estimation of a
critical point by such an extrapolation method seems to be difficult
in the case of nuclear multifragmentation. Sizes of fragmenting systems
created in nuclear reactions are not well controlled
due to the preequilibrium emission and their range is limited.
In addition, one may expect large departures from the scaling relation
of Eq. (1) for very small systems.
Our observation is that, without relying on the finite-size scaling,
the best estimation of the critical point is given by
the position of the maximum of
$\kappa_{2}/\langle s_{max}\rangle ^2$.
Behavior of this quantity for different system sizes and
various event sortings will be investigated in the following sections.

\section{Order parameter fluctuations}

\subsection{Cumulants and finite-size scaling}

The order parameter probability distribution representative
for small lattices with open boundaries is shown in Fig. 2
for various values of $p$.
Far from the transition the distribution is sharply
peaked with an extended tail to the right (left) in the disordered
(ordered) phase and positioned close to the limiting values.
In the transition region the distribution rapidly evolves 
passing through a broad, flattened and (nearly) symmetrical
distribution.
This behavior can be well characterized by using the skewness, $K_{3}$,
measuring the asymmetry of a distribution and the kurtosis excess,
$K_{4}$, which quantifies the degree of peakedness.
\begin{figure}[ht]
\includegraphics[width=3.375in]{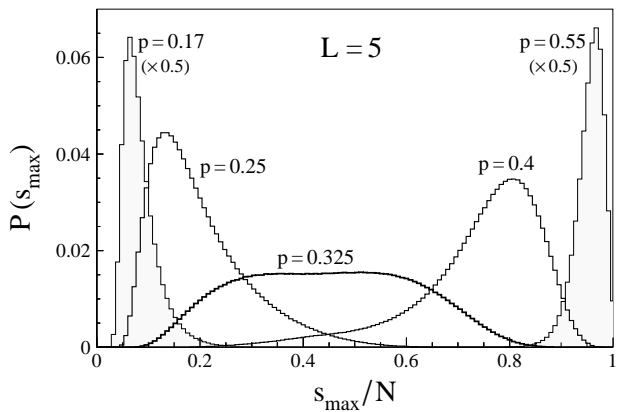}
\caption{Probability distributions of the largest cluster size
for a lattice of size $N=125$ with free boundary conditions.}
\label{fig:2}
\end{figure}
The quantities of interest are defined and
expressed in terms of the cumulants as
\begin{eqnarray}
K_2\equiv &\mu_{2}/\langle s_{max}\rangle^2& =\kappa_2/\kappa_1^2 \nonumber\\
K_3\equiv &\mu_{3}/\mu_{2}^{3/2}& =\kappa_3/\kappa_2^{3/2} \nonumber\\
K_4\equiv &\mu_{4}/\mu_{2}^{2}-3& =\kappa_4/\kappa_2^{2},
\end{eqnarray}
where $\mu_i=\langle(s_{max}-\langle s_{max}\rangle)^i\rangle$ is the $i$-th
central moment, and $\kappa_i$ is the $i$-th cumulant of the $P(s_{max})$
distribution. They are plotted as
a function of $p$ for different system sizes in Fig. 3(a).
In the vicinity of the critical point and for $L\rightarrow\infty$
one expects for these dimensionless parameters the scaling relation
\begin{equation}
K_{i}=f_{i}[(p-p_{c})L^{1/\nu}].
\end{equation}
If the scaling holds, values of $K_{i}$
at $p_c$ are independent of the system size, which is approximately
observed in our plots as the crossing of curves for different $L$ near $p_c$.
\begin{figure}[ht]
\includegraphics[width=3.375in]{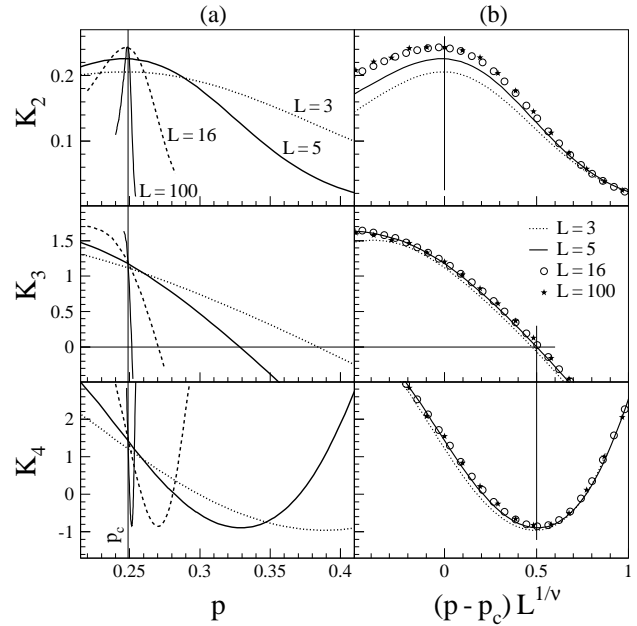}
\caption{The cumulant ratios of Eq. (2) as a function of the bond probability (a)
and the scaling variable (b) for systems of different sizes.
Calculations with free boundary conditions.}
\label{fig:3}
\end{figure}
To verify the scaling in the neighborhood of the critical point
the $K_i$ parameters are replotted in Fig. 3(b) against the
scaling variable $(p-p_{c})L^{1/\nu}$, where $\nu=7/8$.
The collapse of the data shows that the scaling relation with no corrections
for finite-size effects is well satisfied even for such small lattices
with open boundaries.
As can be estimated from Fig. 3(b) the asymptotic values of $K_i$ at $p_c$
are about $K_{2}=0.24$, $K_{3}=1.2$, and $K_{4}=1.5$.
For small systems they are somewhat smaller with
largest deviations observed for $K_2$.
A prominent feature of the $K_2$ distribution is the maximum
located very close to $p_c$ irrespective of the system size.

Another characteristic point, corresponding to the broad transitional
distribution shown in Fig. 2, is where $K_3=0$ and $K_{4}$ reaches
its minimum value of about $-0.9$, and $K_{2}\simeq 0.135$.
It is observed at some distance from the critical point,
which depends on the system size as $(p-p_{c})\simeq 0.5L^{-1/\nu}$.
This point approximately coincides with the maximum of the mean cluster size
and the power-law behavior of the fragment size distribution (see Fig. 4),
and may be used as an estimation of the pseudocritical point.
The line in Fig. 4 is a power-law fit of Eq. (1)
to such points giving $\nu=0.878\pm0.005$
in agreement with the expected value $\nu=0.875$.
Corrections to the scaling are not significant in this case.
Other variables considered in Fig. 4 show
much larger deviations from the asymptotic scaling behavior.

The above characteristics are for free boundary conditions.
Behavior of $K_i$ for periodic boundary conditions
when surface effects are reduced is shown
in Fig. 5. Also in this case the finite-size scaling features
are clearly observed. As expected, the scaling functions $f_i$ are
compressed now toward lower bond probabilities, which can be seen
by comparing Fig. 5(b) with Fig. 3(b).
The pseudocritical point is positioned very close
to the critical point. Thus, the large difference in locations
of these points in the case of free boundary conditions may be
interpreted as a surface effect.
\begin{figure}[ht]
\includegraphics[width=3.375in]{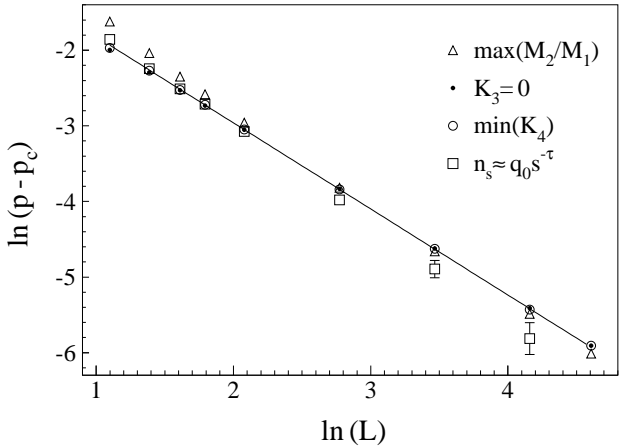}
\caption{Finite-size scaling plot for values of the bond probability, $p$,
at which the conditions indicated on the figure are fulfilled. The line is
the best linear fit to the open circles.}
\label{fig:4}
\end{figure}
\begin{figure}[ht]
\includegraphics[width=3.375in]{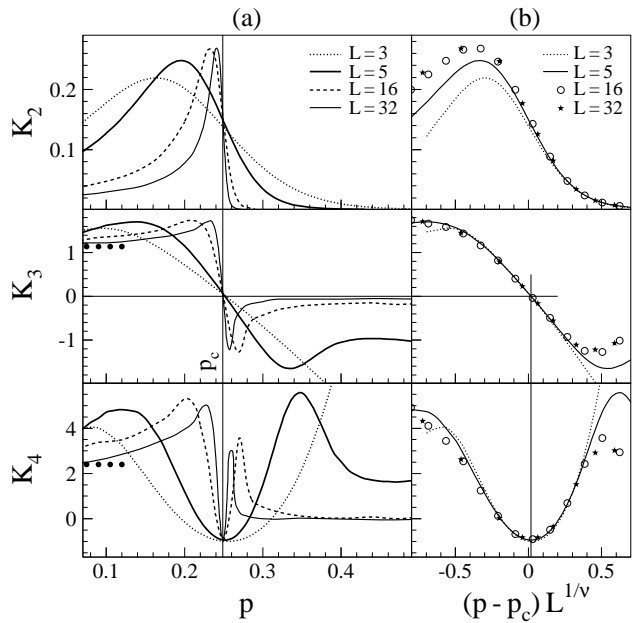}
\caption{Same as Fig. 3 for periodic boundary conditions.
The dots in the left part indicate the $K_3$ and $K_4$ values for
the Gumbel distribution (see text).}
\label{fig:5}
\end{figure}
\begin{figure}[ht]
\includegraphics[width=3.375in]{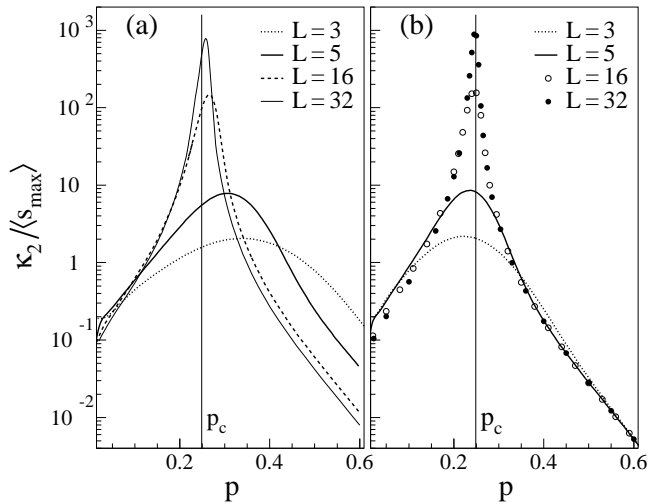}
\caption{The normalized variance of $P(s_{max})$
versus the bond probability for free boundary conditions (a),
and periodic boundary conditions (b).}
\label{fig:6}
\end{figure}
\subsection{Delta scaling}
Behavior of the cumulant moments is of interest in the context
of $\Delta$-scaling proposed for studying criticality
in finite systems \cite{d-p,d-e,uf,d-n}.
Probability distributions $P(s_{max})$ of the extensive
order parameter, $s_{max}$, for different ``system sizes'',
$\langle s_{max}\rangle$, obey $\Delta$-scaling if they
can be converted to a single scaling function
$\Phi(z_{(\Delta)})$ by the transformation
\begin{equation}
\langle s_{max}\rangle^{\Delta}P(s_{max})=\Phi(z_{(\Delta)})
\equiv \Phi \left( \frac{s_{max}-\langle s_{max}\rangle}
{\langle s_{max}\rangle^{\Delta}} \right),
\end{equation}
where $1/2 \leq \Delta \leq 1$.
It was argued that the order parameter satisfies the
``first-scaling law'' ($\Delta=1$ scaling) for systems at the critical
point and in the disordered phase. The ``second-scaling law''
($\Delta=1/2$) applies for systems in the ordered phase
far from the transition point.
Using a three-dimensional bond percolation model, it was demonstrated
that the $\Delta=1$ scaling holds near the critical point
while the $\Delta=1/2$ scaling is satisfied above the percolation
threshold at $p=0.35$ \cite{d-e,uf}.
However, these results were obtained for rather large systems
$N=14^{3}$ to $32^3$ with periodic boundary conditions.
In the following, we examine the scaling properties
in a wide range of the control parameter, also for
smaller systems with open boundaries.

In case of a $\Delta$-scaling the normalized cumulants
$\kappa_{i}/\kappa_{1}^{i\Delta}$ are independent of the
``system size'', $\kappa_{1}\equiv\langle s_{max}\rangle$.
Therefore, for a set of $P(s_{max})$ distributions,
$K_{3}=\mathrm{const}$ and $K_{4}=\mathrm{const}$ are necessary conditions
for a $\Delta$-scaling. Additionally, $K_{2}=\mathrm{const}$ for $\Delta=1$,
and $\kappa_{2}/\kappa_{1}=\mathrm{const}$ for $\Delta=1/2$.
The conditions for the $\Delta=1$ scaling are
fulfilled only in the vicinity of the critical point. The crossing
points seen in Figs. 3(a) and 5(a) indicate
the system size independence of $K_{i}$ near $p_{c}$. The conditions
are also satisfied around the critical point when
both the system size and the control parameter are
varied so that $(p-p_{c})L^{1/\nu}=\mathrm{const}$, accordingly
to the finite-size scaling.
Since the conditions are necessary but not
sufficient, we have checked that indeed the scaling relation
of Eq. (4) is approximately satisfied.
Observing the $\Delta=1$ scaling
requires then a variation of the system size without or with
a very specific change of the control parameter.
It cannot be observed when the system size is fixed.
A similar conclusion has been reached in Ref. \cite{gul}
for a lattice gas model. This contradicts the statement
of Refs. \cite{d-e,uf,d-n} that the scaling relation
is valid independently of any phenomenological reasons
for changing the ``system size''.

Fig. 5(a) shows no evidence for the presence of a $\Delta$-scaling
in the subcritical region $p<p_{c}$ (disordered phase). The largest
cluster size in subcritical percolation have been 
extensively studied in Ref. \cite{baz1}. As predicted by
the theory of extremes of independent
random variables, $P(s_{max})$ converges to the
Fisher-Tippett (Gumbel) distribution when $N\rightarrow\infty$.
The mean grows logarithmically with the system size
while the variance is bounded. Such a behavior cannot be described by
a $\Delta$-scaling.
For the Gumbel distribution $K_{3}\simeq 1.14$ and $K_{4}=2.4$,
marked in Fig. 5(a) by the dots. As can be seen, the small systems
show significant deviations from these asymptotic values
even for periodic boundary conditions.

The limiting behavior of the largest cluster size
in the supercritical region $p>p_{c}$ (ordered phase)
is governed by the Central Limit Theorem \cite{uf,baz2}.
The asymptotic distribution
is Gaussian, $K_{3}=K_{4}=0$, with the mean and variance
both increasing linearly with the system size,
$\kappa_{2}/\kappa_{1}=\mathrm{const}$,
satisfying the $\Delta =1/2$ scaling relation.
Such characteristics are seen in Figs. 5(a) and 6(b)
away from the critical point for larger systems
with periodic boundary conditions. Since the normalized
variance $\kappa_{2}/\kappa_{1}$, same for all $N$,
systematically decreases with $p$, the scaling can only be
observed for fixed $p$ and different $N$.
Considering very small systems $L=3$ to 6 with free
boundary conditions as appropriate for nuclear applications,
Fig. 6(a) shows that $\kappa_{2}/\kappa_{1}$ at fixed $p$
strongly depends on the system size.
This indicates the violation of the scaling as a consequence of
surface effects. 
\begin{figure}[ht]
\includegraphics[width=3.375in]{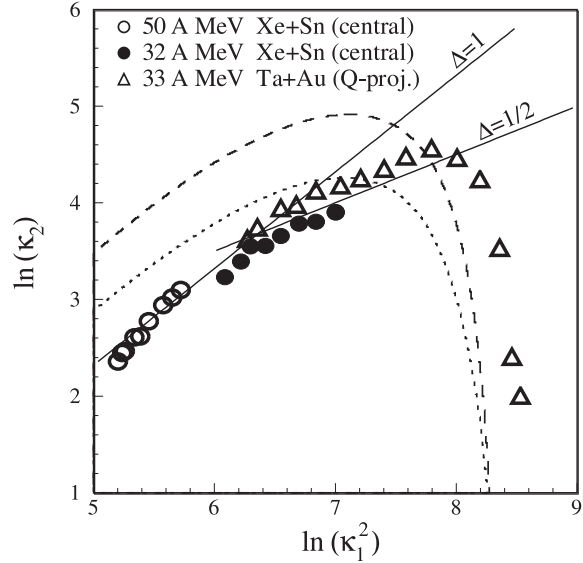}
\caption{Log-log plot of the variance versus the squared mean
value of the largest fragment size (charge) distribution.
The points are experimental data for events sorted
according to the excitation energy \cite{fra}.
The lines are percolation results for $N=64$ when
events are sorted by the bond probability (dashed line)
and by the fraction of open bonds (dotted line).}
\label{fig:7}
\end{figure}
\begin{figure}[ht]
\includegraphics[width=3.375in]{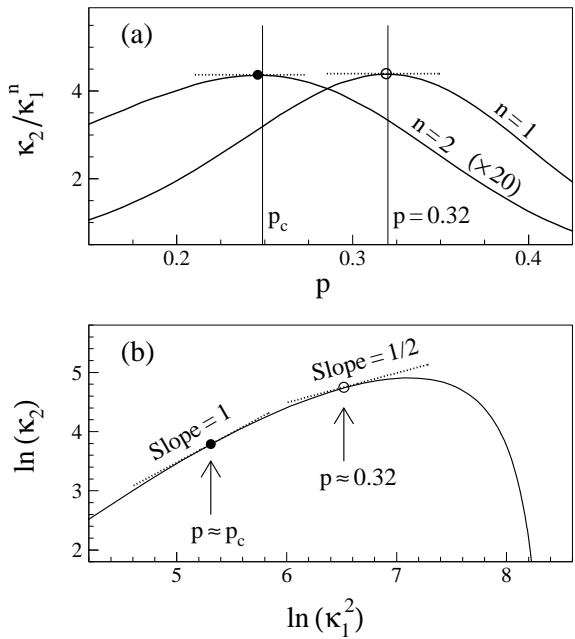}
\caption{Percolation lattice of 64 sites with open boundaries.
(a) The normalized variance of $P(s_{max})$ as a function of the bond
probability. (b) The variance versus the squared mean in log-log representation
for events binned by $p$.}
\label{fig:8}
\end{figure}

Investigations of the largest fragment charge distribution, $P(Z_{max})$,
observed in heavy-ion central collisions at bombarding energies
between 25 and 150 $A$MeV have shown that
$\kappa_{2}/\kappa_{1}\simeq \mathrm{const}$
at lower energies while $K_{2}\simeq \mathrm{const}$
in a high energy range \cite{d-p,fra,frank}.
The two regimes appear on the $\ln(\kappa_{2})$ versus
$\ln(\kappa_{1}^2)$ plot along two lines with the slope of $1/2$ and 1.
This observation has been interpreted as the presence
of the two limiting $\Delta$-scaling laws
corresponding to the ordered and disordered phases (in the case of
$\Delta$-scaling the slope is equal to $\Delta$).
Fig. 7 shows such a plot when experimental events are sorted according
to the estimated source excitation energy.
These data taken from Ref. \cite{fra} include quasi-projectiles from
Ta + Au collisions allowing to observe
a strong suppression of the fluctuations at lowest energies.
Assuming the percolation pattern of the cumulants,
interpretation of the high energy branch in terms of the
$\Delta=1$ scaling would require that for different excitation
energies all fragmenting systems are created with nearly
the same value of a control parameter close to a critical
point. More realistically, the control parameter varies
with the excitation energy while changes of the system size are less
significant. The dashed and dotted lines in Fig. 7 show percolation
results for fixed $N=64$ when events are binned by the
bond probability and by the fraction of open bonds.
Given the mean, the variance depends not only on the system size
but also on the choice of binning variable.
For a quantitative comparison with the data one would have to determine
appropriate system sizes
and a sorting variable equivalent to the excitation energy.
Nevertheless, the qualitative behavior of the lines shows
similarity to the experimental data.
The model suggests that the slope changes continuously and,
as is shown in Fig. 8, the point with slope of 1
corresponds to the maximum of $K_2$ (locally $K_2\simeq \mathrm{const}$),
whereas the slope of $1/2$ reflects the maximum of $\kappa_{2}/\kappa_{1}$.
These features are not related to a $\Delta$-scaling. The rise 
and fall behavior of the correlation in Fig. 7 is a simple consequence
of the mass conservation constraint. Some points on this line may have
a particular meaning depending on the assumed model.
In the present model, the point of slope 1 approximately corresponds
to the critical point. Within the canonical lattice gas model
the point of maximum variance occurs close to the critical point
at the critical density. For subcritical densities it is located
inside the coexistence region \cite{gul}.

\subsection{Event sorting effects}
The cumulant properties of the largest cluster size distribution
presented above are for an ideal situation,
which assumes that generated events are sorted according to precisely
known values of the control parameter. In experimental studies
such a selection is difficult to realize. Usually the sorting parameter
is a measurable quantity (e.g., multiplicity, excitation energy per nucleon)
which is correlated with the control parameter with a significant dispersion.
\begin{figure}[ht]
\includegraphics[width=3.375in]{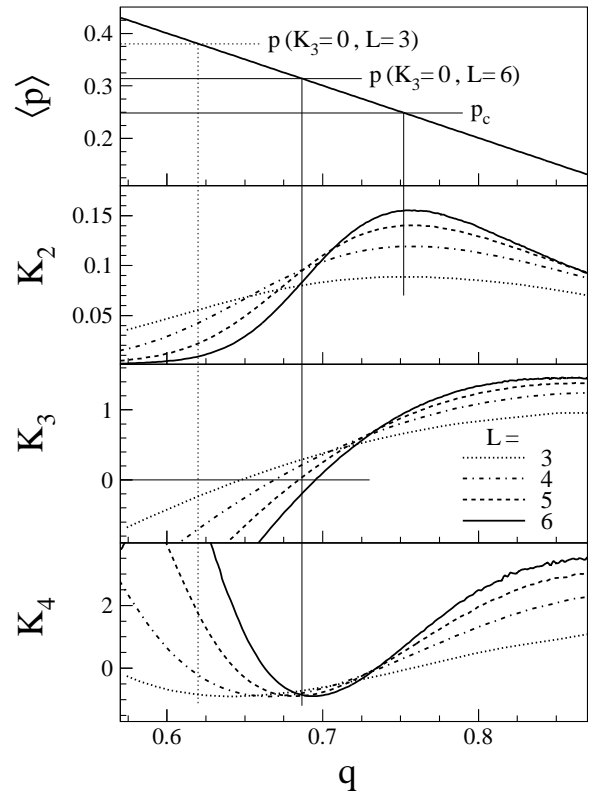}
\caption{The cumulant ratios as a function of the fraction of open bonds;
on top the correspondence to the mean value of the bond probability.}
\label{fig:9}
\end{figure}
Even if an attempt is made to estimate a control parameter such as
the temperature in nuclear multifragmentation, some dispersion is 
unavoidable. This can be simulated by using the fraction of open
bonds, $q$, for sorting events. On average, the relation between
$p$ and $q$ is linear, $q\simeq 1-p$, for all system sizes.
Fig. 9 presents $K_i$ as a function of $q$ for small lattices
of $L=3$ to $6$ with open boundaries which are relevant
for nuclear multifragmentation studies.
The top diagram shows the correspondence between $q$ and $\langle p\rangle$.
Comparing $K_{i}(q)$ with $K_{i}(p)$ in corresponding intervals shows that
absolute values may change significantly, however, some characteristic features
are approximately preserved. The $K_{2}(q)$ distributions exhibit maxima near
the ``critical'' value $q_{c}\simeq0.75$ corresponding to $p_c$. The zeros of
$K_{3}(q)$ coincide with the minima of $K_{4}(q)$ reflecting the behavior near
the pseudocritical point. However, at the ``critical'' point, $q_c$, the
cumulant values are now much smaller and differ with the lattice size.
The crossing points appear shifted from $q_c$ toward the ordered phase.
These intersection points are spread out over some range of $q$ and are
expected to converge to $q_c$ with increasing $N$. The relation between $p$
and the number of broken bonds is governed by a binomial distribution.
This implies that for a given $q$, the dispersion of $p$ vanishes as $\sim N^{-1/2}$
with $N\rightarrow\infty$. On the other hand, according to the finite-size
scaling, $K_i(p)$ follows the same pattern irrespective of $N$ within
a fixed interval of the scaling variable $(p-p_{c})L^{1/\nu}$.
Thus, the corresponding interval of $p$ vanishes with increasing $N$ as
$\sim N^{-1/3\nu}\simeq N^{-1/2.625}$. Since the dispersion vanishes faster,
the limiting distributions $K_i(q)$ and $K_i(p)$ will be equivalent:
$K_i(q)\rightarrow K_i(p=1-q)$ when $N\rightarrow\infty$.
Significant differences between the sortings are observed in small systems.
\begin{figure}[ht]
\includegraphics[width=3.375in]{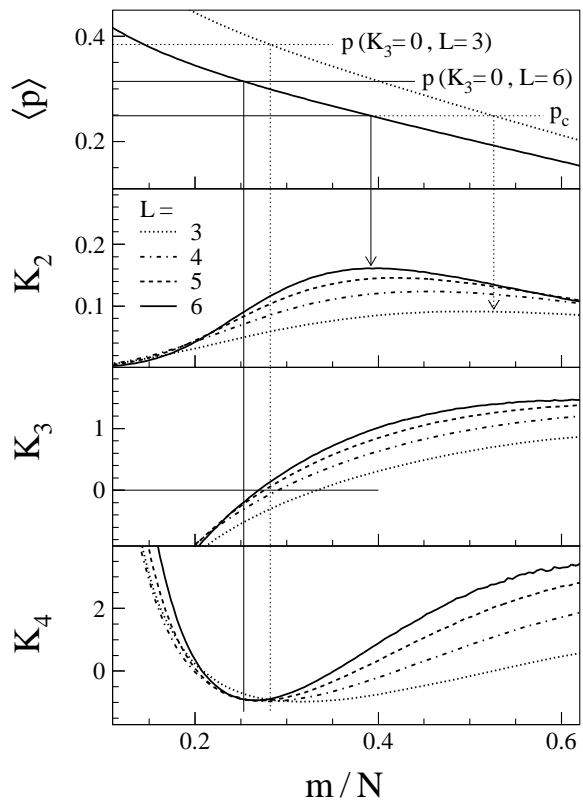}
\caption{The cumulant ratios as a function of the normalized multiplicity.}
\label{fig:10}
\end{figure}
\begin{figure}[ht]
\includegraphics[width=3.375in]{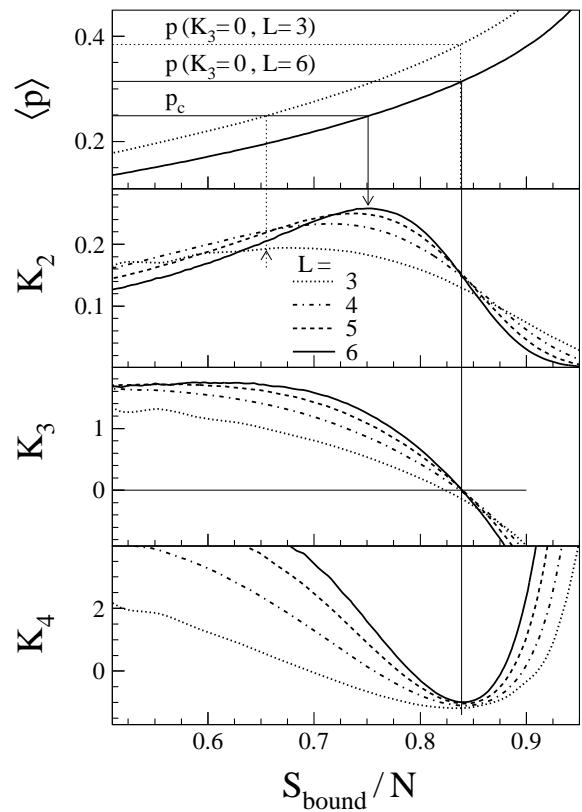}
\caption{The cumulant ratios as a function of the normalized
total size of complex fragments.}
\label{fig:11}
\end{figure}
\begin{figure}[ht]
\includegraphics[width=3.375in]{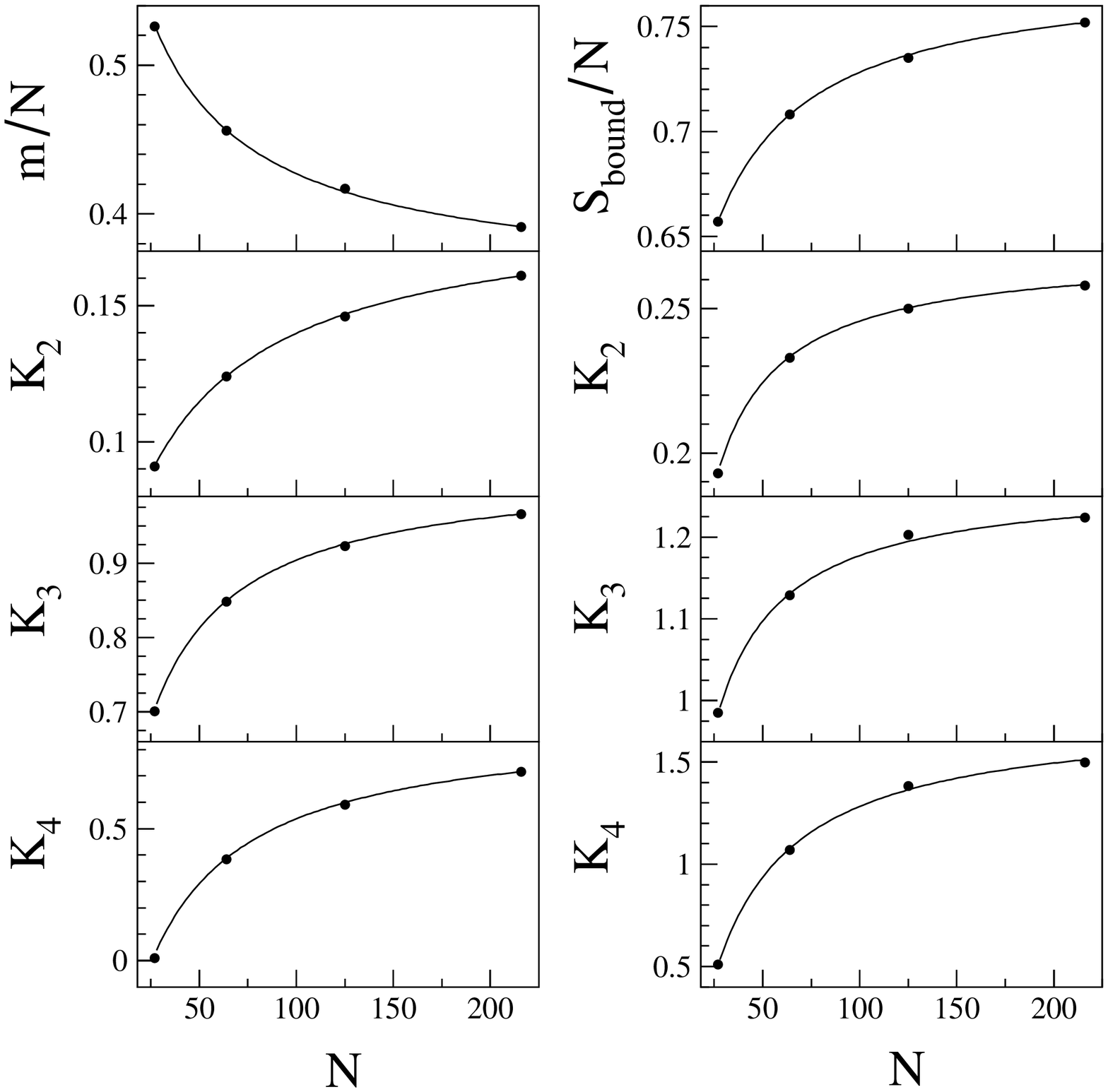}
\caption{Parameters of the ``critical'' point as a function of the system size
when events are binned according to the multiplicity (left column) and
$S_{bound}$ (right column). The points show percolation values, the lines
represent approximations by Eq. (5).}
\label{fig:12}
\end{figure}

Particularly interesting is sorting events according to
directly measurable quantities. We have examined the $K_i$
dependencies on the overall multiplicity, $m$, (Fig. 10)
and the total size of all clusters
of size greater than $1$, $S_{bound}$, (Fig. 11).
The parameters are normalized to the system size $N$. 
In both cases the average correspondence to the control parameter depends on
the system size, which is shown in the top diagrams where the solid
line is for $L=6$ and the dotted one for $L=3$.
The $K_2$ distributions exhibit maxima whose positions correspond to the
critical value of the control parameter, $p_c$.
For such ``near-critical'' events, values of the sorting variables and
the cumulant ratios depend on the system size as shown in Fig. 12.
They all can be well described by the equation
\begin{equation}
X=a-\frac{c}{N+b},
\end{equation}
with coefficients $a,b$, and $c$ listed in Table I.
In practice, when the system size in not well known,
one can examine $K_i$ as a function of $m$ (and/or $S_{bound}$).
The system size $N$ can be found by solving Eq. 5 with $X=m/N$
for $m=m_c$ at which the maximum of $K_2$ is observed.
Then, the values of $K_{2}, K_{3},$ and $K_{4}$ calculated from Eq. 5
can be verified with those observed at $m=m_c$.

For all the considered sortings, $K_{3}=0$ occurs at the same position
as the minimum of $K_4$, near the maximum of the mean cluster size
and a point where the best power-law fit to the fragment size
distribution is observed. Therefore, such a point may be used as an alternative
or complementary indication
of the pseudocritical point. The minimum value of $K_4$ is close to $-1$ for
all the sorting parameters and system sizes.

Figs. 10 and 11 show that the crossing points, as it was for the sorting
by $q$, are shifted from the ``critical'' point corresponding to $p_c$
toward the ordered phase region.
For $S_{bound}/N$ they are well localized near the pseudocritical point.
In this case the pseudocritical point can be additionally characterized
by $K_{2}\simeq 0.15$ and
the position $S_{bound}/N \simeq 0.84$.

In our simulations events have been generated for uniformly distributed
values of the bond probability, $p$, and then grouped in bins of
a sorting variable. Using a different distribution of $p$
changes the spectrum of events in a bin. Consequently,
quantities such as $K_i$ calculated for events within a bin may also
change their values. Calculations performed for a gaussian distribution
of $p$ which might simulate experimental conditions
have shown that this effect is of minor importance.

\begin{table}
\caption{The coefficients of Eq. (5) for events grouped in bins of
the multiplicity (upper part) and $S_{bound}$ (lower part).}
\begin{ruledtabular}
\begin{tabular}{cddd}
   X & a & b & c \\
\hline
$m/N$ & 0.352 & 28. & -9.6 \\
$K_{2}$ & 0.187 & 44. & 6.8 \\
$K_{3}$ & 1.03 & 19. & 15. \\
$K_{4}$ & 0.91 & 26. & 47. \\
\\
$S_{bound}/N$ & 0.777 & 23. & 6. \\
$K_{2}$ & 0.27 & 7. & 2.6 \\
$K_{3}$ & 1.27 & 8. & 10. \\
$K_{4}$ & 1.74 & 16. & 53. \\
\end{tabular}
\end{ruledtabular}
\end{table}

\section{Conclusions}

The largest cluster size distributions have been examined
within a percolation model on small lattices with open
boundaries. The dimensionless cumulant ratios
as the normalized variance, $K_2$, the skewness, $K_3$,
and the kurtosis, $K_4$, of the distribution
satisfy with a good accuracy the finite-size scaling
in the critical region.
In particular, $K_i$ are independent of the system size
near the critical point (crossing points). This feature has been explored
in phase transition studies as the cumulant crossing method,
particularly for the kurtosis in
the form of the Binder cumulant \cite{binder}. To our knowledge
this method has not been applied in analyzing multifragmentation
data. However, it would require a wide range of system sizes, which
is difficult to realize in nuclear reactions.
Moreover, the presence of crossing points has an
unambiguous interpretation when events are sorted according to
the control parameter. In practice, events are grouped with an
inevitable dispersion over the control parameter.
This blurs the scaling behavior; the crossing points
may appear in a wide range of the sorting parameter away from
the ``critical'' point. These remarks apply also for
the $\Delta=1$ scaling law, since the occurrence of the finite-size scaling
for the cumulant ratios is a necessary condition for this scaling law.
The model shows that the $\Delta$-scaling method
fails for systems in normal phases. The largest cluster size fluctuations
in the disordered phase cannot be described by a $\Delta$ scaling,
whereas the limiting $\Delta=1/2$ scaling in the ordered phase
is violated in small systems with open boundaries.

The percolation transition in such systems can be, however,
identified by examining some distinct features of the finite-size
scaling functions of $K_i$,
which are not significantly affected by corrections to scaling in
small systems and by a dispersion of the control parameter when events
are sorted according to various measurable quantities.
The maximum of $K_2$ approximately corresponds
to the location of the true critical point.
The absolute values of $K_i$ at this point have been
determined for sortings by the control parameter, the multiplicity, and
$S_{bound}$, providing complementary characteristics.
Coincidentally with the maximum of the mean cluster size (pseudocritical point)
and the power-law behavior of the fragment size distribution,
one observes $K_{3}=0$ and a minimum value of
$K_{4}\simeq -1$. If the quantity $S_{bound}$ is used for sorting events,
this point can be additionally characterized by $K_{2}\simeq 0.15$,
and its location is related to the system size as $S_{bound}\simeq 0.84N$.
The analysis does not require the knowledge nor variation of
the system size, which is not well controlled in nuclear multifragmentation.
It allows to estimate the system size at the critical and pseudocritical
points.

It will be interesting to confront these predictions
with multifragmentation data, in particular with the Aladin data,
in which the sorting parameter $Z_{bound}$ and the charge of the largest
fragment are well determined in a wide range of the excitation energy.
It would be also instructive to perform similar analysis with other models
which are known to contain or not contain a phase transition or critical
behavior. Using appropriate system sizes, boundary conditions
and event sortings in model simulations is an important requirement.

\smallskip

This work was supported by the Polish Scientific Research Committee,
Grant No. 2P03B11023.

\end{document}